\documentclass[aps,prb,twocolumn,groupedaddress,showpacs]{revtex4}

\usepackage[dvips,dvipdfm]{graphicx}
\usepackage[usenames,dvipsnames]{color}

\begin{document}
\title{Magnetotunneling Between Two-dimensional Electron Gases in InAs-AlSb-GaSb Heterostructures}
\author{Y.~Lin}
\altaffiliation{Now at NTT Basic Research Laboratories, Atsugi, Japan}
\author{E.~M.~Gonz\'{a}lez}
\altaffiliation{Permanent address: Universidad Complutense, Madrid, Spain}
\author{E.~E.~Mendez}
\email[Corresponding author: ]{Emilio.Mendez@sunysb.edu}
\affiliation{Department of Physics and Astronomy, State University of New York at Stony Brook, Stony Brook, NY 11794-3800}
\author{R.~Magno}
\author{B.~R.~Bennett}
\author{A.~S.~Bracker}
\affiliation{Naval Research Laboratory, Washington, DC 20375-5347}
\date{\today}

\begin{abstract}
We have observed that the tunneling magnetoconductance between two-dimensional (2D) electron gases formed at nominally identical InAs-AlSb interfaces most often exhibits two sets of Shubnikov-de Haas oscillations with almost the same frequency. This result is explained quantitatively with a model of the conductance in which the 2D gases have different densities and can tunnel between Landau levels with different quantum indices. When the epitaxial growth conditions of the interfaces are optimized, the zero-bias magnetoconductance shows a single set of oscillations, thus proving that the asymmetry between the two electron gases can be eliminated.
\end{abstract}
\maketitle
The unusual energy-band alignment of heterostructures that combine InAs and GaSb (or their alloys) has made them attractive for electronic devices like interband-tunneling diodes and quantum-cascade lasers. Lately, those structures have also emerged as one of the leading candidates for ``spin" devices, in which the electron's spin states rather than its charge determine the operation of the device, be it a transistor~\cite{Datta90} or a resonant-tunneling diode.~\cite{Koga02a}  

In a III-V compound-semiconductor heterostructure with an asymmetric potential profile, the two spin subbands of the conduction band are split for non-zero in-plane wave vectors in the absence of a magnetic field. This splitting, frequently called Rashba splitting,~\cite{Rashba60,Bychkov84} is particularly large (several meV) in InAs because its band-structure parameters favor a large spin-orbit-coupling coefficient, and may lead to two populations of electrons with opposite spin orientations.

On the long road to practical spin devices, one of the first steps has been to establish unequivocally the presence of Rashba splitting in a given heterostructure, to determine its amount, and, if possible, to control it. The effect has manifested itself as a beating pattern in the Shubnikov-de Haas (SdH) oscillations of the in-plane magnetoresistance.~\cite{Luo88,Das90} Sometimes, an external electric field perpendicular to the layers has been used to further modify the relative carrier densities.~\cite{Nitta97} However, some experimental results on the Rashba splitting have been inconclusive. For instance, Brosig \emph{et~al.} did not observe any beating in the SdH oscillations of asymmetric InAs-AlSb quantum wells with various carrier densities;~\cite{Brosig99} and Heida \emph{et~al.} found no significant effect on the beating pattern by a gate voltage.~\cite{Heida98} Moreover, the presence of a beating pattern in magnetoresistance oscillations may signal a phenomenon quite different from spin splitting, for example  mixing of the first-subband series with magneto-interband oscillations.~\cite{Rowe01}  

Since some of the proposed spin devices rely on electron tunneling across a heterostructure, it is equally important to sort out effects in vertical transport that might be construed as evidence of spin splitting.~\cite{Yamada02} Recently we observed a beating pattern in the zero-bias tunneling magnetoconductance of an InAs-AlSb-GaSb-AlSb-InAs heterostructure with a potential profile adequate for a spin-filter device.~\cite{Gonzalez01} However, that pattern was explained not as a consequence of Rashba splitting but in terms of tunneling between two-dimensional electron gases (2DEGs) with slightly different densities.~\cite{Gonzalez01} This conclusion was based on an analysis that used a very simple magnetotunneling model, in which only one Landau level at a time participated in the tunneling process.    

In this paper, we report a systematic magnetotunneling study using a variety of samples with different thickness for the GaSb well and the AlSb barriers, and employing a more elaborate transport model, which allows for tunneling involving several (energy-broadened) Landau levels simultaneously and in which the Landau-level index may not be conserved in the process. This study not only confirms the generality of our initial conclusion but also shows how the epitaxial growth of the heterostructures affects dramatically the beating pattern, which disappears when the materials interfaces are optimized. 

The band profile common to the InAs-AlSb-GaSb-AlSb-InAs heterostructures discussed here is depicted in the central inset of Fig. 1, and in greater detail in Ref.~\onlinecite{Gonzalez01}. Since the top of the valence band of GaSb is higher in energy than the bottom of the InAs conduction band, two-dimensional (2D) electrons accumulate in quasi-triangular wells formed at each of the two InAs interfaces. Ideally, the number of 2D electrons in each interface should be the same, and the sum of the two equal to the number of 2D holes left in the central GaSb valence-band well. 

To illustrate the similarities and differences among the many structures we have studied, we focus on four heterostructures prepared by molecular beam epitaxy (MBE) at different institutions and in a time span of almost a decade: sample A, with GaSb and AlSb thicknesses of 75 {\AA} and 25 {\AA}, respectively; sample B, with corresponding 60 {\AA} and 34 {\AA} widths; and samples C and D, both with 82 {\AA} and 31 {\AA} GaSb and AlSb regions, respectively.~\cite{growth} Samples A and B were grown a few weeks apart but several years earlier than samples C and D, which were grown on the same day, and as explained later, differed from each other only on the procedure followed to prepare one of the InAs-AlSb interfaces. 
\begin{figure}[!h]
\includegraphics[width=3.25in,clip=]{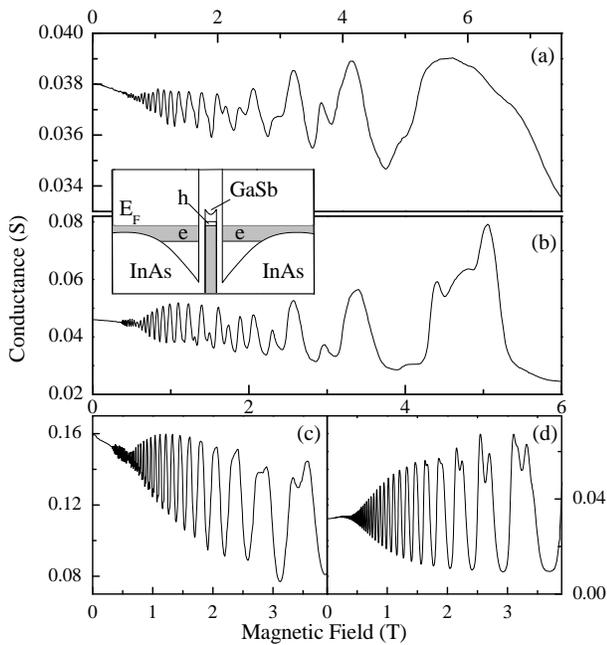}
\caption{\label{fig:SdHs}Shubnikov -- de Haas oscillations of various InAs-AlSb-GaSb-AlSb-InAs heterostructures described in the text: (a) sample A at T=4.2 K, (b) sample B at T=1.7 K, (c) sample C and (d) sample D at T=1.7K. The insert illustrates the potential profile common to all samples, showing the InAs conduction band, the GaSb valence band and AlSb (not listed) in between. 2D electron (e) and hole (h) gases reside in InAs and GaSb layers, respectively. $E_{F}$ is the Fermi energy.}
\end{figure}

The central panel, (b), of Fig.~\ref{fig:SdHs} shows the zero-bias tunneling conductance versus magnetic field (perpendicular to the layers) for sample B, the focus of our earlier study.~\cite{Gonzalez01} As discussed  then, the magnetoconductance reveals SdH oscillations with a beating-like pattern between 0.5 T and 1.5 T, followed by doublet structures (unrelated to spin splitting) with varying relative intensities up to about 4 T. The conductance of sample A [top panel, Fig. 2(a)] exhibits a behavior qualitatively similar to that of sample B, although the characteristic fields and relative intensities of the doublet oscillations are somewhat different. 

In sample C [Fig.~\ref{fig:SdHs}(c)] there are also a beating-like pattern and a doublet region, but they are ``squeezed'' to a narrow field interval, between 0.4 T and 1.2 T, so some of the features are extremely sharp and easy to miss. Above 1.2 T a simple set of oscillations with a single period develops, and Landau-level spin splitting becomes visible at 3.5 T. Finally, the oscillatory behavior of the conductance of sample D [Fig.~\ref{fig:SdHs}(d)] is simple throughout the entire field range, without any trace of beats or doublets (aside from spin splitting above 2 T). 

A fast-Fourier-transform analysis of the conductance oscillations revealed the main frequency component, but failed to yield clear evidence of more than one component, except for sample B. This is not too surprising, in view of the poor beating pattern observed, particularly for samples A and C. When analyzed with the help of a simple model for the tunneling magnetoconductance, the beating and doublet structures found in sample B were attributed to small differences in the carrier densities of the 2D electrons at the InAs interfaces.~\cite{Gonzalez01} In the following, we describe a similar approach to analyze the results of samples A to C using a model that keeps the essence of the previous one but is more general in the treatment of tunneling between Landau levels. 

The basic assumption underlying our model is that the oscillations in the conductance observed in Fig.~\ref{fig:SdHs} are directly related to magnetic states of electrons in the InAs regions only. This assignment is based on the temperature dependence of the amplitude of those oscillations, from which an effective mass is extracted that agrees with the electronic effective mass in InAs. In our model we account for the Landau quantization of 2D electrons in both InAs accumulation regions but ignore the Landau quantization of holes in the GaSb quantum well. The hole gas is treated simply as a ``window'' through which electrons tunnel from one accumulation layer to the other. This simplification is reasonable, considering the small cyclotron energy of heavy holes (when compared to that of electrons in InAs) and the fact that we have not found any features in the experimental magnetoconductance that could be attributed to Landau-quantized holes.

With those premises, we model the conductance as due to tunneling between 2D electron gases (in general, with different electron density) separated by a potential barrier, first assuming that in-plane momentum is conserved throughout the process. We follow Lyo's treatment, in which the tunneling conductance is expressed as~\cite{Lyo98}
\begin{equation}
\label{eq:G}
G = {{4\pi e^{2}g}\over {\hbar}}J_{0}^{2} \sum_{n_{1}, n_{2}}\, \delta_{n_{1},n_{2}} \times \int_{-\infty}^{\infty}[-f'(\zeta )]\rho_{1n}(\zeta )\rho_{2n}(\zeta )\, d\zeta\, .
\end{equation}
$J_{0}$ is the zero-field tunneling integral, $g$ is the Landau level degeneracy per spin, and $n_{i}$ is the Landau index for the $i_{th}$ layer. $ f' (\zeta )$ is the energy derivative of the Fermi-Dirac distribution function at energy $\zeta $. The term $\delta_{n_{2}, n_{1}}$ (= 1 when $n_{2} = n_{1}$ and 0 otherwise) reflects the conservation of in-plane momentum in the presence of a perpendicular magnetic field. $\rho_{in}$ is the density of states (DOS) of the $n_{th}$ Landau level in the $i_{th}$ layer ($i=1,2$). To approximate Eq. (\ref{eq:G}), we use a Gaussian density of states
\begin{equation}
\rho_{in}(\zeta ) = {1 \over {\sqrt{2\pi}\Gamma}}\, exp\, \Big( -{{(\zeta - \varepsilon_{in})^{2}}\over {2\Gamma^{2}}} \Big)\, ,
\end{equation}
with a half-width broadening
\begin{equation}
\Gamma = \frac{1}{2}\frac{\eta\, \hbar\, \omega_{c}}{H^{1/2}}\, ,
\end{equation}
where $\varepsilon_{in}$ is the energy of the $n_{th}$ Landau level in the $i_{th}$ layer, $\eta$ is a materials-dependent constant, $\omega_{c}$ is the cyclotron frequency, and $H$ is the magnetic field perpendicular to the interfaces. 

The chemical potential, $\mu $, is determined by keeping the total carrier density constant~\cite{Zawadzki84}
\begin{equation}
\label{eq:N}
N_{s} = A \sum_{n} \sqrt{\frac{2}{\pi}} \frac{1}{\gamma} \int_{0}^{\infty} \frac{1}{1+ e^{z-\mu '}}\, e^{-2 y_{n}^{2}}\, dz,
\end{equation}
with
\[ y_{n} = \frac{1}{\gamma} (z - \frac{\varepsilon_{n}}{k_{B}T}), \]
and $A=eH/\pi \hbar$, $\gamma = 2\Gamma /k_{B}T$, $z = \zeta/k_{B}T$, and $\mu' = \mu /k_{B}T$.

In the zero-temperature limit, Eq. (\ref{eq:G}) gets simplified to
\begin{equation}
G = {{4\pi e^{2}g}\over {\hbar}}J_{0}^{2} \sum_{n}\rho_{1n}(\mu ) \rho_{2n}(\mu - \Delta)\, ,
\label{eq:GHz}
\end{equation}
where $\Delta = \varepsilon _{2}- \epsilon_{1}$ is the energy difference between the two subbands at $H=0$, and therefore is proportional to the 2D carrier difference. Eq. (\ref{eq:N}) is also simplified accordingly.
\begin{figure}[!tp]
\centering
\includegraphics[width=3.25in,clip=]{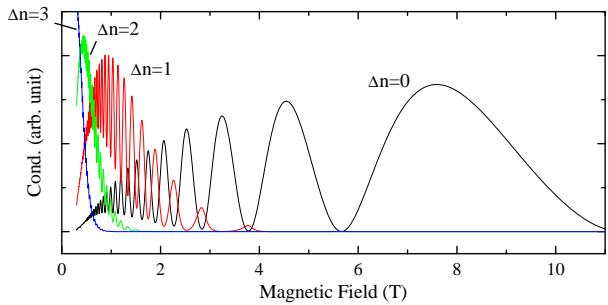}
\caption{\label{fig:CalCond}Various contributions to the magnetoconductance from tunneling channels whose initial and final Landau levels have indices that differ by $\Delta n$ from 0 to 3, calculated using the parameters for sample B (described in the text).}
\end{figure}

With these T = 0 K equations, we can easily calculate the zero-bias tunneling conductance between two 2D electron gases with different carrier densities. The main trace in Fig.~\ref{fig:CalCond} (curve $\Delta n=0$) is the result of such a calculation, when the following parameters (corresponding to sample B) are used: m$_{e} = 0.027$ m$_{0}$, $\Delta = 4$ meV, $N_{s} = 5.7\times 10^{11}$ cm$^{-2}$, and $\eta$ = 0.8. As seen in Fig.~\ref{fig:CalCond}, even though $\Delta \neq 0$ the calculated magnetoconductance exhibits only one set of oscillations, contrary to the experimental result of Fig.~\ref{fig:SdHs}(b). The single set in the calculation is understandable: in Eq. (\ref{eq:GHz}) the product of two density of states for Landau levels whose centers are shifted by an energy $\Delta$ but have the same index (and therefore the same rate of change with field) still looks just like one single density of states.~\cite{errata} 

\begin{figure}[!h]
\centering
\includegraphics[height=3in,angle=90,clip=]{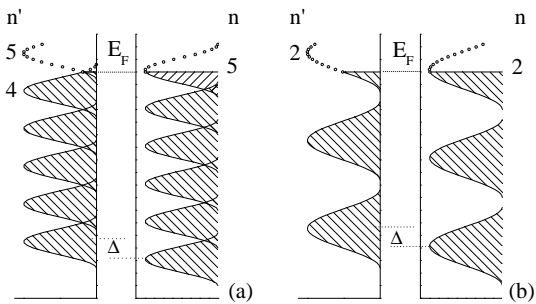}
\caption{\label{fig:illustrate}An illustration of tunneling betwen Landau levels in which the Landau index may not be conserved. Two 2DEGs with subband-energy difference $\Delta$ are separated by a barrier. E$_{F}$ is the Fermi energy. $n$, $n'$ are the Landau indices of the right and the left 2DEGs, respectively. The shadowed regions correspond to filled states. Electrons of Landau index $n$ tunnel into (a) $n'=n$ and  $n$ -1, or (b) the same index, $n$, whenever the Fermi energy falls into the DOS of that Landau level.}
\end{figure}

To explain the existence of more than one set of oscillations, in addition to different carrier densities in the 2D accumulation layers one needs to include the possibility of tunneling between Landau levels with different quantum numbers, in other words, to relax the in-plane momentum conservation law. Let us consider emitter and collector accumulation layers whose zero-field subband energies differ by an amount $\Delta$ as a result of different carrier densities. At low magnetic fields, that is, when $\Delta$ is larger than or comparable to the level broadening $\Gamma$ and to the cyclotron energy $\hbar \omega_{c}$, the Landau levels in one electrode are significantly misaligned with respect to those in the other electrode. Then, for a fixed field the dominant contribution to the density of states at the Fermi level will correspond to one Landau level for the emitter and to another for the collector [see Fig.~\ref{fig:illustrate}(a)]. If we now allow for tunneling transitions in which $\Delta n \neq 0$ more tunneling ``channels'' become open in Eq. (\ref{eq:G}).

On the other hand, at high fields, when $\hbar \omega_{c}$ and $\Gamma$ are much larger than $\Delta$, the misalignment between Landau levels with the same index in the two electrodes becomes relatively insignificant [see Fig.~\ref{fig:illustrate}(b)]. In other words, the contribution of the $\Delta n \neq 0$ channels to the conductance is important at low fields but diminishes rapidly as the field increases. The situation is illustrated in Fig.~\ref{fig:CalCond}, where, in addition to the $\Delta n=0$ channel, we show the individual contributions to the calculated conductance of the $\Delta n=1-3$ channels.

In order to account quantitatively for the experimental dependence of the conductance on magnetic field, it is important to keep in mind that the contributions of the various $\Delta n$ channels to the conductance may not all be the same. Their relative weight will depend on details of the scattering mechanisms responsible for the violation of the $\Delta n=0$ rule. In the absence of a theory that includes those details, we have determined that weight empirically from a fit of the experimental conductance to an appropriately modified Eq. (\ref{eq:G}). 

\begin{figure}[!h]
\center
\includegraphics[width=3.25in,clip=]{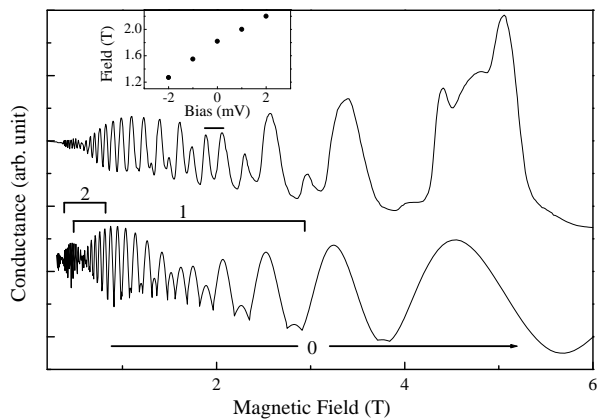}
\caption{\label{fig:Cal}A calculation (lower curve) for the tunneling conductance that includes several $\Delta n$ channels reproduces well the features seen in the experimental result (upper curve) for sample B. The numbers 0, 1, and 2 correspond to the difference between the indices of initial and final Landau levels, $\Delta n$. A short line above the upper curve around 1.8 T indicates the change of the dominant contribution to tunneling, from $\Delta n=1$ to $\Delta n=0$. The insert shows the field dependence of this change when a small bias is applied to the sample.}
\end{figure}

Figure~\ref{fig:Cal} compares the experimental and calculated conductance for sample B, using adequate probability ratios for $\Delta n=0$, 1 and 2. A good fit is obtained when the probability is proportional to $(1+\mid \Delta n\mid )^{-1/2}$. As seen in the figure, for fields approximately below 2 T the $\Delta n=2$ and $\Delta n=1$ transitions are successively dominant. At about 2 T, the $\Delta n=1$ and $\Delta n=0$ transitions have comparable strength and their corresponding oscillations similar height, as highlighted by the short line over the experimental curve. Above 3 T the conductance is almost completely governed by the $\Delta n=0$ transition. 

The ``cross-over" magnetic field (about 1.8 T in Fig.~\ref{fig:Cal}) at which the $\Delta n=1$ and $\Delta n=0$ contributions to the conductance are comparable is closely related to the difference between the subbands energies of the two 2D electron gases, or, equivalently, to the carrier asymmetry between the two gases. This asymmetry is affected by a small bias between the electrodes, which slightly changes the potential profile along the tunneling direction and the carrier density in the accumulation layers but does not alter the near-equilibrium condition (insignificant tunneling current) of this study. In the insert of Fig.~\ref{fig:Cal} we show the value of the cross-over field for several biases, from -2 mV to +2 mV. With increasing positive bias (of the top electrode relative to the substrate electrode) the cross-over field increases, indicating an increasing asymmetry. On the other hand, with negative bias the field becomes smaller and the asymmetry decreases.

A fit of the calculated conductance to the experimental one yields the individual carrier densities in the 2D gases. The difference between the two densities, $\Delta N_{e}$, and their average value, $N_{ave}$, are listed in Table~\ref{table}, along with the subband energy difference $\Delta$. A look at this table reveals the sensitivity of our analysis, which permits us to discern differences of less than 2 percent (sample C) in the two carrier populations. We can therefore say with confidence that in sample D, for which no double oscillations were observed even at the lowest fields (0.3 T), the two gases have essentially the same electron density. 

What the above analysis does not explain is the physical origin of the carrier asymmetry and its difference from one sample to another. Could it be due to the Rashba effect, or is it caused by some external factor, for example the heterostructure growth process? Although it cannot be ruled out completely, it is unlikely that the former effect is the cause of the asymmetry summarized in Table I. The Rashba splitting is closely related to the electric field along the direction of the heterostructure's potential profile, which in turn is proportional to the 2D electron density in the accumulation potential. If the Rashba splitting were behind the observed asymmetry, then for structures of comparable carrier densities, such as those in Table I, one would expect similar values for the parameter $\Delta$. In sharp contrast, $\Delta$ ranges from 10.6 meV to 0 meV.

\begin{table}
\caption{\label{table}Average carrier density, $N_{ave}$, difference between the carrier densities of the two gases, $\Delta N_{e}$, and energy difference between their two ground subbands, $\Delta$ (using m$_{e}=0.027$ m$_{0}$), for samples A to D. $N_{ave}$ was determined from a direct analysis of the SdH oscillations, while $\Delta N_{e}$ was obtained from a detailed comparison between the theoretical and experimental magnetoconductance oscillations.~\cite{CalvsExp}}
\vspace{0.2cm}
\begin{ruledtabular}
\begin{tabular}{ccccc}
& A& B & C & D\\
\hline
N$_{ave}$($10^{11}$ cm$^{-2}$) & 6.2 & 5.0 & 7.0 & 6.5\\
$\Delta N_{e}$($10^{11}$ cm$^{-2}$) & 1.2 & 0.5 & 0.1 & 0\\
$\Delta$ (meV) &10.6 & 4.4 & 0.9 & 0\\
\end{tabular}
\end{ruledtabular}
\end{table}

A far more likely explanation for the carrier asymmetry lies in the epitaxial growth process itself. Though the heterostructures were nominally symmetric with respect to the central GaSb layer, the actual structures might be slightly asymmetric. For instance, it is well known that dopants diffuse along the growth front. It is then possible for the effective thickness of the undoped InAs region near the substrate to be smaller than that near the surface of the structure, with the corresponding asymmetry in the accumulation-layer profile. A small unintentional difference in thickness for the AlSb barriers can also result in a difference in the amount of charge transferred to each of the InAs accumulation layers. Finally, the quality of the material interfaces might affect the number of interface states, which in InAs-AlSb-GaSb heterostructures are known to be a source of extrinsic charges. 

This last possibility is vividly illustrated by comparing the behavior of samples C and D, which nominally have identical layer thicknesses. Their current-voltage characteristics were practically identical to each other from T = 300 K to 4.2 K, exhibiting large negative differential conductance even at room temperature. However, while sample C showed two sets of magnetoconductance oscillations [see Fig.~\ref{fig:SdHs}(c)] and had $\Delta$ = 0.9 meV (see Table I), sample D showed a single set [Fig.~\ref{fig:SdHs}(d)] and $\Delta$ = 0. In addition, in sample D the spin splitting was well resolved at a lower field. 

Samples C and D were grown one immediately after the other and under the same conditions, with one exception. The growth of both samples was optimized to form an InSb-type bond at the AlSb-InAs interfaces, which has been shown to produce a smoother interface than an AlAs-type bond.~\cite{Nosho99b,Tahraoui00} However, sample D was prepared by adding an extra 1/4 In monolayer on the bottom AlSb-InAs interface. This extra layer reduces interface roughness even further, as revealed by plan-view scanning tunneling microscopy.~\cite{Nosho99b,Magno01}

Since the observation of a double set of oscillations in the magnetoconductance requires both an asymmetry in the carrier density and a violation of the $\Delta n=0$ condition, in principle it could be argued that the effect of the added smoothness at the interfaces of sample D is not to eliminate that asymmetry but just to reduce scattering processes that destroy Landau-level-index conservation. In fact, the reduction of scattering in sample D is evident in the appearance of ShH oscillations and spin splitting states at lower fields than for sample C. 

To discern between these possibilities we studied the tunneling magnetoconductance while a small bias was applied between the electrodes. Such bias is not expected to change significantly the scattering processes but it can affect the potential asymmetry, as we have seen above for sample B. Under a 5-meV bias, the magnetoconductance of sample D showed a clear double-oscillation behavior at low fields, indicative of two different electron populations. We can therefore conclude that the exceptional smoothness achieved in sample D leads to identical carrier densities in the two 2D electron gases. 

In summary, we have shown that the presence of beating and double oscillations in the tunneling magnetoconductance between 2D electron gases in nominally symmetric heterostructures is due to the difference in the carrier densities of the two gases. The origin of this difference lies in an extrinsic asymmetry that can be eliminated by reducing interface roughness during the epitaxial growth process. In the samples with the smoothest interfaces the conductance showed a single set of oscillations from the lowest magnetic fields, spin splitting was resolved above 2 T, and beyond 10 T pronounced features were observed corresponding to fractional occupation of the magnetic levels.~\cite{Mendez02} On the one hand, these results cast some doubts on the apparent strength of the Rashba splitting and its suitability for spin devices based on tunneling. On the other, the results are encouraging in that they show that, under optimum growth conditions, balanced populations of high-mobility 2D gases can be formed in InAs accumulation layers, with which 2D-2D interaction effects can be probed.

We are grateful to W. I. Wang, who provided some of the samples used in this study, and to B. Z. Nosho, L. J. Whitman, and B. V. Shanabrook for their assistance in developing MBE techniques applied to the growth of samples C and D discussed here. This work has been supported by the US Army Research Office and the Office of Naval Research.

\begin{thebibliography}{21}
\expandafter\ifx\csname natexlab\endcsname\relax\def\natexlab#1{#1}\fi
\expandafter\ifx\csname bibnamefont\endcsname\relax
  \def\bibnamefont#1{#1}\fi
\expandafter\ifx\csname bibfnamefont\endcsname\relax
  \def\bibfnamefont#1{#1}\fi
\expandafter\ifx\csname citenamefont\endcsname\relax
  \def\citenamefont#1{#1}\fi
\expandafter\ifx\csname url\endcsname\relax
  \def\url#1{\texttt{#1}}\fi
\expandafter\ifx\csname urlprefix\endcsname\relax\def\urlprefix{URL }\fi
\providecommand{\bibinfo}[2]{#2}
\providecommand{\eprint}[2][]{\url{#2}}

\bibitem[{\citenamefont{Datta and Das}(1990)}]{Datta90}
\bibinfo{author}{\bibfnamefont{S.}~\bibnamefont{Datta}} \bibnamefont{and}
  \bibinfo{author}{\bibfnamefont{B.}~\bibnamefont{Das}},
  \bibinfo{journal}{Appl. Phys. Lett.} \textbf{\bibinfo{volume}{56}},
  \bibinfo{pages}{665} (\bibinfo{year}{1990}).

\bibitem[{\citenamefont{Koga et~al.}(2002)\citenamefont{Koga, Nitta,
  Takayanagi, and Datta}}]{Koga02a}
\bibinfo{author}{\bibfnamefont{T.}~\bibnamefont{Koga}},
  \bibinfo{author}{\bibfnamefont{J.}~\bibnamefont{Nitta}},
  \bibinfo{author}{\bibfnamefont{H.}~\bibnamefont{Takayanagi}},
  \bibnamefont{and} \bibinfo{author}{\bibfnamefont{S.}~\bibnamefont{Datta}},
  \bibinfo{journal}{Phys. Rev. Lett.} \textbf{\bibinfo{volume}{88}},
  \bibinfo{pages}{126601} (\bibinfo{year}{2002}).

\bibitem[{\citenamefont{Rashba}(1960)}]{Rashba60}
\bibinfo{author}{\bibfnamefont{E.~I.} \bibnamefont{Rashba}},
  \bibinfo{journal}{Sov. Phys. Solid-State} \textbf{\bibinfo{volume}{2}},
  \bibinfo{pages}{1109} (\bibinfo{year}{1960}).

\bibitem[{\citenamefont{Bychkov and Rashba}(1984)}]{Bychkov84}
\bibinfo{author}{\bibfnamefont{Y.~A.} \bibnamefont{Bychkov}} \bibnamefont{and}
  \bibinfo{author}{\bibfnamefont{E.~I.} \bibnamefont{Rashba}},
  \bibinfo{journal}{J. Phys. C} \textbf{\bibinfo{volume}{17}},
  \bibinfo{pages}{6039} (\bibinfo{year}{1984}).

\bibitem[{\citenamefont{Luo et~al.}(1988)\citenamefont{Luo, Munekata, Fang, and
  Stiles}}]{Luo88}
\bibinfo{author}{\bibfnamefont{J.}~\bibnamefont{Luo}},
  \bibinfo{author}{\bibfnamefont{H.}~\bibnamefont{Munekata}},
  \bibinfo{author}{\bibfnamefont{F.~F.} \bibnamefont{Fang}}, \bibnamefont{and}
  \bibinfo{author}{\bibfnamefont{P.~J.} \bibnamefont{Stiles}},
  \bibinfo{journal}{Phys. Rev. B} \textbf{\bibinfo{volume}{38}},
  \bibinfo{pages}{10142} (\bibinfo{year}{1988}).

\bibitem[{\citenamefont{Das et~al.}(1990)\citenamefont{Das, Datta, and
  Refenberger}}]{Das90}
\bibinfo{author}{\bibfnamefont{B.}~\bibnamefont{Das}},
  \bibinfo{author}{\bibfnamefont{S.}~\bibnamefont{Datta}}, \bibnamefont{and}
  \bibinfo{author}{\bibfnamefont{R.}~\bibnamefont{Refenberger}},
  \bibinfo{journal}{Phys. Rev. B} \textbf{\bibinfo{volume}{41}},
  \bibinfo{pages}{8278} (\bibinfo{year}{1990}).

\bibitem[{\citenamefont{Nitta et~al.}(1997)\citenamefont{Nitta, Akazaki,
  Takayanagi, and Enoki}}]{Nitta97}
\bibinfo{author}{\bibfnamefont{J.}~\bibnamefont{Nitta}},
  \bibinfo{author}{\bibfnamefont{T.}~\bibnamefont{Akazaki}},
  \bibinfo{author}{\bibfnamefont{H.}~\bibnamefont{Takayanagi}},
  \bibnamefont{and} \bibinfo{author}{\bibfnamefont{T.}~\bibnamefont{Enoki}},
  \bibinfo{journal}{Phys. Rev. Lett.} \textbf{\bibinfo{volume}{78}},
  \bibinfo{pages}{1335} (\bibinfo{year}{1997}).

\bibitem[{\citenamefont{Brosig et~al.}(1999)\citenamefont{Brosig, Ensslin,
  Warburton, Nguyen, Brar, Thomas, and Kroemer}}]{Brosig99}
\bibinfo{author}{\bibfnamefont{S.}~\bibnamefont{Brosig}},
  \bibinfo{author}{\bibfnamefont{K.}~\bibnamefont{Ensslin}},
  \bibinfo{author}{\bibfnamefont{R.~J.} \bibnamefont{Warburton}},
  \bibinfo{author}{\bibfnamefont{C.}~\bibnamefont{Nguyen}},
  \bibinfo{author}{\bibfnamefont{B.}~\bibnamefont{Brar}},
  \bibinfo{author}{\bibfnamefont{M.}~\bibnamefont{Thomas}}, \bibnamefont{and}
  \bibinfo{author}{\bibfnamefont{H.}~\bibnamefont{Kroemer}},
  \bibinfo{journal}{Phys. Rev. B} \textbf{\bibinfo{volume}{60}},
  \bibinfo{pages}{R13989} (\bibinfo{year}{1999}).

\bibitem[{\citenamefont{Heida et~al.}(1998)\citenamefont{Heida, {van Wees},
  Kuipers, Klapwijk, and Borghs}}]{Heida98}
\bibinfo{author}{\bibfnamefont{J.~P.} \bibnamefont{Heida}},
  \bibinfo{author}{\bibfnamefont{B.~J.} \bibnamefont{{van Wees}}},
  \bibinfo{author}{\bibfnamefont{J.~J.} \bibnamefont{Kuipers}},
  \bibinfo{author}{\bibfnamefont{T.~M.} \bibnamefont{Klapwijk}},
  \bibnamefont{and} \bibinfo{author}{\bibfnamefont{G.}~\bibnamefont{Borghs}},
  \bibinfo{journal}{Phys. Rev. B} \textbf{\bibinfo{volume}{57}},
  \bibinfo{pages}{11911} (\bibinfo{year}{1998}).

\bibitem[{\citenamefont{Rowe et~al.}(2001)\citenamefont{Rowe, Nehls, Stradling,
  and Ferguson}}]{Rowe01}
\bibinfo{author}{\bibfnamefont{A.~C.~H.} \bibnamefont{Rowe}},
  \bibinfo{author}{\bibfnamefont{J.}~\bibnamefont{Nehls}},
  \bibinfo{author}{\bibfnamefont{R.~A.} \bibnamefont{Stradling}},
  \bibnamefont{and} \bibinfo{author}{\bibfnamefont{R.~S.}
  \bibnamefont{Ferguson}}, \bibinfo{journal}{Phys. Rev. B}
  \textbf{\bibinfo{volume}{63}}, \bibinfo{pages}{201307}
  (\bibinfo{year}{2001}).

\bibitem[{\citenamefont{Yamada et~al.}(2002)\citenamefont{Yamada, Kikutani,
  Gozu, Sato, and Kita}}]{Yamada02}
\bibinfo{author}{\bibfnamefont{S.}~\bibnamefont{Yamada}},
  \bibinfo{author}{\bibfnamefont{T.}~\bibnamefont{Kikutani}},
  \bibinfo{author}{\bibfnamefont{S.}~\bibnamefont{Gozu}},
  \bibinfo{author}{\bibfnamefont{Y.}~\bibnamefont{Sato}}, \bibnamefont{and}
  \bibinfo{author}{\bibfnamefont{T.}~\bibnamefont{Kita}},
  \bibinfo{journal}{Physica E} \textbf{\bibinfo{volume}{13}},
  \bibinfo{pages}{815} (\bibinfo{year}{2002}).

\bibitem[{\citenamefont{Gonz{\'a}lez et~al.}(2001)\citenamefont{Gonz{\'a}lez,
  Lin, and Mendez}}]{Gonzalez01}
\bibinfo{author}{\bibfnamefont{E.~M.} \bibnamefont{Gonz{\'a}lez}},
  \bibinfo{author}{\bibfnamefont{Y.}~\bibnamefont{Lin}}, \bibnamefont{and}
  \bibinfo{author}{\bibfnamefont{E.~E.} \bibnamefont{Mendez}},
  \bibinfo{journal}{Phys. Rev. B} \textbf{\bibinfo{volume}{63}},
  \bibinfo{pages}{033308} (\bibinfo{year}{2001}).

\bibitem[{gro()}]{growth}
\bibinfo{note}{Typical growth conditions can be found in
  Ref.\onlinecite{Gonzalez01} for Sample A and B, and in
  Ref.\onlinecite{Magno01} for Sample C and D (380 $^{o}$C growth
  temperature).}

\bibitem[{\citenamefont{Lyo}(1998)}]{Lyo98}
\bibinfo{author}{\bibfnamefont{S.~K.} \bibnamefont{Lyo}},
  \bibinfo{journal}{Phys. Rev. B} \textbf{\bibinfo{volume}{57}},
  \bibinfo{pages}{9114} (\bibinfo{year}{1998}).

\bibitem[{\citenamefont{Zawadzki and Lassnig}(1984)}]{Zawadzki84}
\bibinfo{author}{\bibfnamefont{W.}~\bibnamefont{Zawadzki}} \bibnamefont{and}
  \bibinfo{author}{\bibfnamefont{R.}~\bibnamefont{Lassnig}},
  \bibinfo{journal}{Surf. Sci.} \textbf{\bibinfo{volume}{142}},
  \bibinfo{pages}{225} (\bibinfo{year}{1984}).

\bibitem[{err()}]{errata}
\bibinfo{note}{Although not mentioned explicitly there, the calculation in
  Ref.~\onlinecite{Gonzalez01}  allowed for the Landau-level indices 
  in the emitter and collector to be different but involved just one Landau level in each electrode.}

\bibitem[{Cal()}]{CalvsExp}
\bibinfo{note}{The value of $N_{s}$ obtained from a fit of Eq. (\ref{eq:GHz})
  to the experimental conductance throughout the entire range of magnetic
  fields is about 10{\%} larger than that determined by a Fourier analysis of
  the SdH oscillations, which is dominated by the more numerous low-field
  oscillations. One possible reason for this small discrepancy is that the
  model assumes constant effective mass and constant carrier density, which in
  reality may not be the case. Another possibility is a small effect due to the
  sample holder, which is slightly magnetic.}

\bibitem[{\citenamefont{Nosho et~al.}(1999)\citenamefont{Nosho, Weinberg,
  Barvosa-Carter, Bracker, Magno, Bennett, Culbertson, Shanabrook, and
  Whitman}}]{Nosho99b}
\bibinfo{author}{\bibfnamefont{B.~Z.} \bibnamefont{Nosho}},
  \bibinfo{author}{\bibfnamefont{W.~H.} \bibnamefont{Weinberg}},
  \bibinfo{author}{\bibfnamefont{W.}~\bibnamefont{Barvosa-Carter}},
  \bibinfo{author}{\bibfnamefont{A.~S.} \bibnamefont{Bracker}},
  \bibinfo{author}{\bibfnamefont{R.}~\bibnamefont{Magno}},
  \bibinfo{author}{\bibfnamefont{B.~R.} \bibnamefont{Bennett}},
  \bibinfo{author}{\bibfnamefont{J.~C.} \bibnamefont{Culbertson}},
  \bibinfo{author}{\bibfnamefont{B.~V.} \bibnamefont{Shanabrook}},
  \bibnamefont{and} \bibinfo{author}{\bibfnamefont{L.~J.}
  \bibnamefont{Whitman}}, \bibinfo{journal}{J. Vac. Sci. Technol. B}
  \textbf{\bibinfo{volume}{17}}, \bibinfo{pages}{1786} (\bibinfo{year}{1999}).

\bibitem[{\citenamefont{Tahraoui et~al.}(2000)\citenamefont{Tahraoui, Tomasini,
  Lassabatere, and Bonnet}}]{Tahraoui00}
\bibinfo{author}{\bibfnamefont{A.}~\bibnamefont{Tahraoui}},
  \bibinfo{author}{\bibfnamefont{P.}~\bibnamefont{Tomasini}},
  \bibinfo{author}{\bibfnamefont{L.}~\bibnamefont{Lassabatere}},
  \bibnamefont{and} \bibinfo{author}{\bibfnamefont{J.}~\bibnamefont{Bonnet}},
  \bibinfo{journal}{Appl. Suf. Sci.} \textbf{\bibinfo{volume}{162}},
  \bibinfo{pages}{425} (\bibinfo{year}{2000}).

\bibitem[{\citenamefont{Magno et~al.}(2001)\citenamefont{Magno, Bracker,
  Bennett, Nosho, and Whitman}}]{Magno01}
\bibinfo{author}{\bibfnamefont{R.}~\bibnamefont{Magno}},
  \bibinfo{author}{\bibfnamefont{A.~S.} \bibnamefont{Bracker}},
  \bibinfo{author}{\bibfnamefont{B.~R.} \bibnamefont{Bennett}},
  \bibinfo{author}{\bibfnamefont{B.~Z.} \bibnamefont{Nosho}}, \bibnamefont{and}
  \bibinfo{author}{\bibfnamefont{L.~J.} \bibnamefont{Whitman}},
  \bibinfo{journal}{J. Appl. Phys.} \textbf{\bibinfo{volume}{90}},
  \bibinfo{pages}{6177} (\bibinfo{year}{2001}).

\bibitem[{\citenamefont{Mendez et~al.}()\citenamefont{Mendez, Lin, Magno, and
  Bennett}}]{Mendez02}
\bibinfo{author}{\bibfnamefont{E.~E.} \bibnamefont{Mendez}},
  \bibinfo{author}{\bibfnamefont{Y.}~\bibnamefont{Lin}},
  \bibinfo{author}{\bibfnamefont{R.}~\bibnamefont{Magno}}, \bibnamefont{and}
  \bibinfo{author}{\bibfnamefont{B.~R.} \bibnamefont{Bennett}},
  \bibinfo{note}{unpublished.}

\end{thebibliography}
\end{document}